\documentclass[floatfix,groupedaddress,twocolumn,showpacs,amsmath,pra,aps,amssymb,longbibliography,nofootinbib]{revtex4-1}

\usepackage{graphicx}
\usepackage{xcolor}
\usepackage{lipsum}
\usepackage{overpic}
\usepackage[english]{babel}
\usepackage{hyperref}
\usepackage{amsmath,amsfonts,amssymb,verbatim,float,mathtools}
\usepackage{physics}
\usepackage[titletoc,title]{appendix} 
\usepackage{soul}
\usepackage{mathtools}

\hypersetup{
    unicode=true, % non-Latin characters in Acrobat?s bookmarks
    plainpages=false,
    colorlinks=true,% false: boxed links; true: colored links
    linkcolor=blue,% color of internal links
    citecolor=blue,% color of links to bibliography
    filecolor=blue,% color of file links
    urlcolor=blue% color of external links
}

\usepackage{bbold}

\newcommand{\e}{\mathcal{E}}

\newcommand{\cbE}{\boldsymbol{\mathbf{\cal E}}}

\renewcommand{\vec}[1]{\mathbf{#1}}

\newcommand\widthscale{1.0}

\begin{document}		
	\title{Signatures of optical phase transitions in super- and subradiant arrays of atoms}
	
	\author{C. D. Parmee}
	\affiliation{Department of Physics, Lancaster University, Lancaster, LA1 4YB, United Kingdom}
	\author{J. Ruostekoski}
	\affiliation{Department of Physics, Lancaster University, Lancaster, LA1 4YB, United Kingdom}
	\begin{abstract}
Resonant light interacting with matter can support different phases of a polarizable medium, and optical bistability where two such phases coexist. Here we identify signatures of optical phase transitions and optical bistability mapped onto scattered light in planar arrays of cold atoms. Methods on how to explore such systems in superradiant, and extreme subradiant states existing outside the light cone, are proposed. The cooperativity threshold and intensity regimes for the intrinsic optical bistability, supported by resonant dipole-dipole interactions alone, are derived in several cases of interest analytically. Subradiant states require lower intensities, but stronger cooperativity for the existence of non-trivial phases than superradiant states. The transmitted light reveals the onset of phase transitions and bistability that are predicted by mean-field theory as large jumps in coherent and incoherent signals and hysteresis. In the quantum solution, traces of phase transitions are identified in enhanced quantum fluctuations of excited level populations. 
	\end{abstract}
	\date{\today} 
	\maketitle

%%%%%%%%%%%%%%%%%%%%%%%%%%%%%%%%%%%%%%%%%%%%%%%%%%%%%%%%%%%%%%%%%%%%%%%%%
\section{Introduction}

Resonant emitters in regular planar arrays have attracted considerable attention from classical circuit resonators forming metamaterials~\cite{Zheludev12} and metasurfaces~\cite{Yu14} to plasmonics~\cite{Halas11} and quantum systems, such as superconducting SQUID rings~\cite{Anlageprx} and cold atoms~\cite{rui2020}. Such surfaces can be utilised for manipulation of electromagnetic fields, including phase-holography~\cite{Arbabi15} 
and sensing~\cite{Tittl18}. In systems where the radiative interactions between closely-spaced emitters are particularly strong, the entire array has been driven to a giant subradiant state~\cite{Jenkins17,rui2020}.
In arrays of closely-spaced cold atoms, the strong light-mediated 
dipole-dipole interactions arise naturally, as atoms do not absorb light,
their resonances are well defined, and the atoms can respond to light quantum-mechanically.
Atomic arrays have been proposed as constituents of metamaterials~\cite{Jenkins2012a}, for quantum information processing~\cite{Grankin18,Guimond2019,ballantine2020}, atomic clocks~\cite{Kramer2016,Henriet2018,Qu19}, emission of nonclassical light~\cite{Olmos16,williamson2020b,Cidrim20} and entanglement~\cite{Ritsch_subr,bettles2019,Asenjo-Garcia19}, and a way to realise topological phases~\cite{Perczel2017b,Bettles2017}. 

Most experiments on collective optical responses of cold atoms so far, in both random atomic ensembles~\cite{BalikEtAl2013,CHA14,Pellegrino2014a,wilkowski,Jennewein_trans,Ye2016,Jenkins_thermshift,vdStraten16,Guerin_subr16,Machluf2018,Dalibard_slab,Bettles18} and in arrays~\cite{rui2020} and chains~\cite{Glicenstein20} with unit occupancy, have focused on the limit of low light intensity (LLI), where the full quantum model can, under appropriate conditions, be reduced to a linear system of $N$ harmonic oscillators \cite{Lee16}. Beyond the LLI regime with multiple excitations, atomic arrays start experiencing saturation, and the rich phenomenology of long-range interactions and collective behaviour can lead to the full many-body quantum solutions deviating from the semiclassical models that neglect quantum fluctuations~\cite{bettles2019}. The differences between quantum and classical solutions in nonlinear systems are widely studied in the context of phase transitions, and in optics, one of the best-known phase transitions is optical bistability~\cite{Lugiato1984} in atomic systems. 

Optical bistability and phase transitions have been actively studied 
in systems without the spatial correlations and structure of the sample~\cite{Gibbs1976,bonifacio1976,Bonifacio1978,Carmichael1977,Agrawal79,Carmichael1986a,Drummond1980,Drummond1981,Savage1988a}, e.g., in cavities where the feedback mechanism is provided by the cavity mirrors~\cite{Rosenberger1983,Orozco1984,Rempe1991}. Intrinsic bistability is
a process where phase transitions are generated by the self-interactions of the sample, and despite having been observed in highly-excited Rydberg atoms in the microwave regime~\cite{Carr2013}, intrinsic bistability was for a long time considered unachievable for atoms with light-mediated interactions. 
Recent theoretical studies that also take into account the spatial structure of the many-body systems suggest that intrinsic bistability and phase transitions are more generic and could occur in a variety of systems with short- and long-range interactions~\cite{Lee2011,Sibalic2016,Parmee2018}. 

For optical systems, it is natural to ask what are the observable signatures of phase transitions and optical bistability, and how these are mapped onto the scattered light.
In this paper, by studying light emission from radiatively strongly coupled atoms in planar arrays of subwavelength spacing, we identify optical signatures of phase transitions
in collective atomic excitations. To do so, we employ periodic boundary conditions and a mean-field approximation which closes the spectral gap in the system, representing a decohered quantum state where the correlations are absent, and compare our results to the full quantum model. 

We develop a simple analytic theory for an intrinsic optical bistability due to radiative interactions between atoms in planar arrays and derive the cooperativity parameter, indicating a bistability threshold $k a < (\pi/3)^{1/2}$, with the lattice spacing $a$ 
and resonance wavenumber $k$. We find that multiple mean-field-theoretical stable phases, including ones with spontaneous symmetry breaking and persistent oscillations, and optical bistability are identifiable in the transmitted light as large jumps in coherent and incoherent signals and hysteresis upon sweeping of the laser frequency.
If the corresponding changes in dipole amplitudes are small, the signal of phase transitions and hysteresis in the coherent transmission is sometimes much weaker than in the incoherent photon count, which still provides sharp peaks, e.g., when moving into regions of antiferromagnetic and oscillatory phases. 
In the quantum solution, the phase transitions and bistability are absent, but traces of them are revealed in enhanced quantum fluctuations of incoherently scattered light, and most clearly in those of the excited level populations.

We find that the response sensitively depends on the underlying LLI collective excitation eigenmode of the corresponding linear system that is targeted by incident light. Bistability can even exist between subradiant and superradiant
modes, providing a method for also preparing subradiant excitations via a laser frequency sweep. For subradiant modes, bistabilities occur at lower intensities 
and the existence of phase transitions requires smaller lattice spacings, $ka\alt 0.34\pi$, compared to the one for a superradiant mode, $ka\alt 0.44\pi$.
We propose methods on how to drive such eigenmodes by manipulating the atomic level shifts and consider two examples: a uniform mode that was recently experimentally studied in subradiant transmission measurements~\cite{rui2020}, which at smaller lattice spacings, considered here,  becomes superradiant, and an extreme subradiant checkerboard eigenmode that can exist outside the light cone, decoupled from the environment.

%%%%%%%%%%%%%%%%%%%%%%%%%%%%%%%%%%%%%%%%%%%%%%%%%%%%%%%%%%%%%%%%%%%%%%%%%%%%
\section{Model}
\label{sec:setup}
	
\subsection{Quantum system of atoms and light}
	
\begin{figure}
	\hspace*{-0 cm}
	\includegraphics[width=\columnwidth]{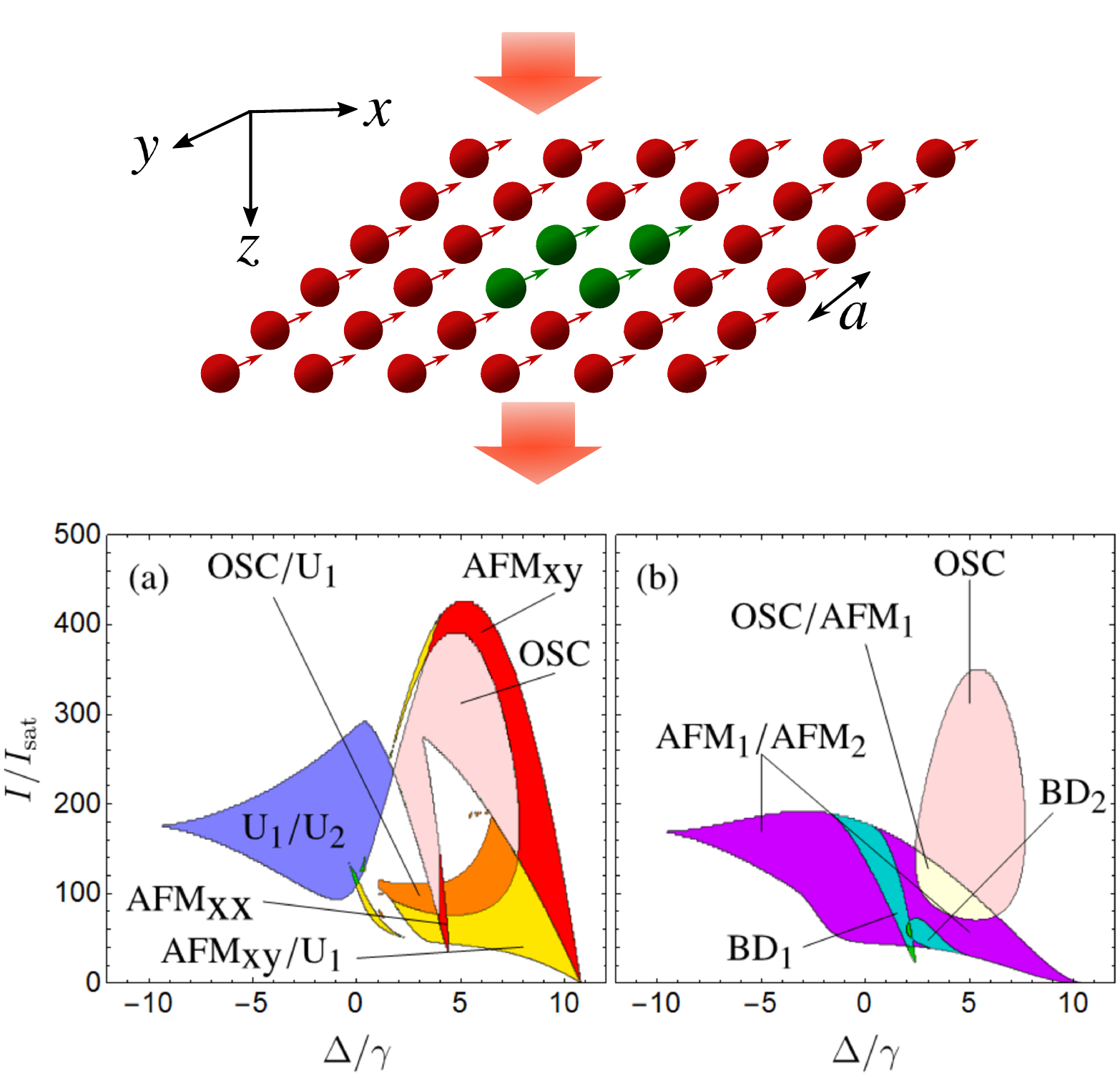}
	\caption{A 2D array of atoms illuminated by incident light and resulting stable phases of atoms as a function of laser frequency and incident light intensity. Top: Only the central four atoms (green) are independent, while the remaining array is obtained by periodic boundary conditions. Bottom: (a) Uniform (b) alternating profiles of level shifts (the different phases are explained in the text).}
	\vspace{-0 cm}
	\label{qxqyplotBP}
\end{figure}

We consider a two-level system of cold atoms trapped in a two-dimensional (2D) array with one atom per site, illuminated by an incident plane wave $\cbE^+(r)=\e_0 \hat{\vec{e}} \exp(ikz)$; Fig.~\ref{qxqyplotBP}. 
We take the polarisation and the direction of the atomic dipoles to be $\hat{\textbf{e}}=-(\hat{\textbf{x}}+\hat{\textbf{y}})/\sqrt{2}$ along the diagonal of the lattice. Light-induced resonant dipole-dipole interactions mediate strong interactions between the atoms. The atomic systems are subject to periodic boundary conditions to simulate an infinite lattice and, for simplicity, we vary the parameters of only $N=4$ atoms in a square array. We also assume that the atoms are sufficiently tightly confined, such that the spatial fluctuations can be neglected. The standard many-body quantum master equation for the atoms in the rotating-wave approximation for slowly varying amplitudes reads~\cite{Lehmberg1970}
\begin{equation}\label{MasterEq}
\begin{split}	\frac{d\hat{\rho}}{dt}&=-\frac{\text{i}}{\hbar}\left[\hat{H}{}-\sum_{jl(j\neq l)}\hbar \Omega_{jl}\hat{\sigma}{}_j^+\hat{\sigma}{}_l^-,\hat{\rho}\right]+\\
&\sum_{jl}\gamma_{jl}\left(2\hat{\sigma}{}_j^-\hat{\rho} \hat{\sigma}{}_l^+- \hat{\sigma}{}_l^+\hat{\sigma}{}^-_j\hat{\rho}-\hat{\rho}\hat{\sigma}{}_l^+\hat{\sigma}{}^-_j\right),	\end{split}
\end{equation}
where the square brackets represent a commutator  and $\hat{\sigma}_j^{+}=|e\rangle_{j}\mbox{}_{j}\langle g|=(\hat{\sigma}_j^-)^{\dagger}$ the raising operator, where $\ket{e}_j$ and $\ket{g}_j$ are the excited and ground state of the two-level atom on site $j$, respectively. The dispersive and dissipative parts of the light-induced dipole-dipole interaction terms are $\Omega_{jl}$ and $\gamma_{jl}$, respectively (see Appendix A for details).
The Hamiltonian is given by
\begin{equation}
\hat{H}=-\sum_{l}\left[\textbf{d}_{eg}\cdot\boldsymbol{\mathcal{E}}^+(\textbf{r}_l)\hat{\sigma}^+_l+\textbf{d}_{ge}\cdot\boldsymbol{\mathcal{E}}^-(\textbf{r}_l)\hat{\sigma}^-_l+\hbar\Delta_l\hat{\sigma}^{ee}_l\right],
\end{equation} 
where $\hat{\sigma}^{ee}_l=\hat{\sigma}_l^{+}\hat{\sigma}_l^{-}$, $\Delta_l=\omega-\omega^{(l)}_{eg}$ is the detuning, $\omega=kc$ the laser frequency, $\omega^{(l)}_{eg}$ the transition frequency of an atom on site $l$, and
$\textbf{d}_{eg}$ the dipole matrix element, with $\textbf{d}_{ge}=\textbf{d}_{eg}^*$. 
We express the incident light intensity $I=2\epsilon_0c |\mathcal{E}_0|^2$ in units of the saturation intensity, $I_{\rm sat}=\hbar c 4\pi^2 \gamma/3\lambda^3$, or the Rabi frequency, ${\cal R}_l=\textbf{d}_{eg}\cdot\boldsymbol{\mathcal{E}}^+(\textbf{r}_l)/\hbar$, as $I/I_{\rm sat}=2({\cal R}/\gamma)^2$, where the single-atom linewidth $\gamma=|\textbf{d}_{eg}|^2 k^3/(6\pi\epsilon_0\hbar)$.
\\

%

%%%%%%%%%%%%%%%%%%%%%%%%%%%%%%%%%%%%%
\subsection{Mean-field approximation}

In addition to the full quantum many-body dynamics, we also consider the Gutzwiller mean-field approximation, $\hat{\rho}\approx \otimes\hat{\rho}_i$, where quantum fluctuations between the atoms are neglected. This corresponds to the factorization of internal level correlations,
$
\langle \hat{\sigma}{}_i^{\alpha}\hat{\sigma}{}_j^{\beta}\rangle\approx \langle \hat{\sigma}{}_i^{\alpha}\rangle\langle \hat{\sigma}{}_j^{\beta}\rangle 
$ ($\alpha \neq \beta$),
since we assume atoms are at fixed positions with no 
%position 
spatial fluctuations, and therefore there are no light-induced correlations~\cite{Lee16,bettles2019} between the atoms after the factorization.
The dynamics then obey the nonlinear equations
\begin{subequations}\label{SpinEqs}
	\begin{align}
	\begin{split}
	\dot{\rho}^{(l)}_{ge}=&\left(\text{i}\Delta_l-\gamma\right) {\rho}^{(l)}_{ge}-\\
	&\text{i}(2{\rho}^{(l)}_{ee}-1)\big[{\cal R}_l+\sum_{j\neq l}(\Omega_{jl}+\text{i}\gamma_{jl}){\rho}^{(j)}_{ge}\big],
	\end{split}\\
	\begin{split}
	\dot{\rho}^{(l)}_{ee}=&-2\gamma {\rho}^{(l)}_{ee}+2\text{Im}[{\cal R}_l^*{\rho}^{(l)}_{ge}]+\\
	&2\text{Im}\big[\sum_{j\neq l}(\Omega_{jl}-\text{i}\gamma_{jl}){\rho}^{(l)}_{ge}({\rho}^{(j)}_{ge})^*\big],
	\end{split}
	\end{align}
\end{subequations}
where $\rho_{ge}^{(l)}=\text{Tr}\{\hat{\sigma}{}^{-}_l\hat{\rho}(t)\}$ and $\rho_{ee}^{(l)}=\text{Tr}\{\hat{\sigma}{}^{ee}_l\hat{\rho}(t)\}$. We will use Eqs.~\eqref{SpinEqs} to determine the long-time phases and optical bistability that can occur in the system.

%%%%%%%%%%%%%%%%%%%%%%%%%%%%
\subsection{Scattered light}

The total light amplitude is the sum of the incident and scattered fields
$\hat{\textbf{E}}{}^{\pm}(\textbf{r})=\boldsymbol{\mathcal{E}}{}^{\pm}(\textbf{r})+\hat{\textbf{E}}{}_s^{\pm}(\textbf{r})$, with the scattered electric field given by the sum of the contributions from all the atoms
\begin{equation}
\begin{split}
\epsilon_0&\hat{\textbf{E}}{}^+_s(\textbf{r})=\sum_{l}\mathsf{G}(\textbf{r}-\textbf{r}_l)\textbf{d}_{ge}\hat{\sigma}{}_l^-,
\end{split}
\end{equation}
where $\mathsf{G}(\textbf{r}-\textbf{r}_l)$ is the dipole radiation kernel~\cite{Jackson} [Eq.~\eqref{dipolekernel} in Appendix A].
We will compare the optical responses obtained from the full quantum dynamics of Eq.~\eqref{MasterEq} with those calculated from the mean-field equations, Eqs.~\eqref{SpinEqs}.
We consider coherently transmitted light in the forward direction, $T_{\rm coh}=|\hat{\textbf{e}}\cdot\langle\hat{\textbf{E}}{}^-(\textbf{r})\rangle|^2/\left|\hat{\textbf{e}}\cdot\boldsymbol{\mathcal{E}}{}^{-}(\textbf{r}) \right|^2$,
expressed in terms of the optical depth $\text{OD}=-\log(T_{\rm coh})$.
We also calculate the rate of scattered photons 
\begin{equation}\label{TransmissionCoh-2}
n=\frac{2\epsilon_0c}{\hbar \omega_0 }\int \langle\hat{\textbf{E}}{}_{s}^-(\textbf{r})\cdot\hat{\textbf{E}}{}^+_{s}(\textbf{r})\rangle dS,
\end{equation}
where $\langle\hat{\textbf{E}}{}_{s}^-(\textbf{r})\cdot\hat{\textbf{E}}{}^+_{s}(\textbf{r})\rangle=\langle\hat{\textbf{E}}{}_{s}^-(\textbf{r})\rangle\cdot\langle\hat{\textbf{E}}{}^+_{s}(\textbf{r})\rangle$ for the coherent and  $\langle\hat{\textbf{E}}{}_{s}^-(\textbf{r})\cdot\hat{\textbf{E}}{}^+_{s}(\textbf{r})\rangle-\langle\hat{\textbf{E}}{}_s^-(\textbf{r})\rangle\cdot\langle\hat{\textbf{E}}{}^+_s(\textbf{r})\rangle$ for the incoherent photon count-rate (ICR) [Eq.~\eqref{IncoherentPhotonCountRateFull} in Appendix A].
The ICR that we will use under the mean-field description [Eq.~\eqref{IncoherentPhotonCountRate} in Appendix A] is different from the usual semiclassical approximation for the incoherent scattering. The atom-light dynamics is solved from the mean-field equations, Eqs.~\eqref{SpinEqs}, but the single-atom quantum description of emitted light $\langle \hat{\sigma}_l^{ee}\rangle$ is now included~\cite{bettles2019} for the scattered light, and the ICR no longer vanishes for atoms at fixed positions~\cite{meystre1998}.

%%%%%%%%%%%%%%%%%%%%%%%%%%%%%%%%%%%%%%%%%%%%%%%%%%%%%%%%%%%%%%%%%%%%%%%%%

\section{Optical signatures of phase transitions and bistability}

%%%%%%%%%%%%%%%%%%%%%%%%%%%%%%%%%%%%%%%%%%%%%%%%%%%%%%%%%%%%%%%%%%%%%
\subsection{Collective  low light intensity eigenmodes}

In the limit of LLI, $\rho_{ee}^{(l)} = 0$, and the mean-field equations, Eqs.~\eqref{SpinEqs}, coincide with the coupled-dipole model of classical linear oscillators driven by light. In this regime we may analyse the optical response using LLI collective radiative excitation eigenmodes and the complex eigenvalues, which represent the collective line shifts (from the resonance of an isolated atom)  $\delta_{{\bf q}}$ and linewidths $\upsilon_{{\bf q}}$. Collective modes with $\upsilon_{{\bf q}}>\gamma$ ($\upsilon_{{\bf q}}<\gamma$) are termed superradiant (subradiant).
Modes with $|\textbf{q}| a > 0.2\pi$ exist outside the light cone and are completely dark in an infinite lattice. 

We focus on two LLI eigenmodes (see Appendix B): the spatially uniform superradiant mode $v_{\rm un}({\bf r}_l)$ ($\upsilon_{\rm uni}=25\gamma$) and a subradiant mode with a checkerboard-patterned phase variation with every atom oscillating $\pi$ out-of-phase from its nearest-neighbour $v_{\rm cb}({\bf r}_l)$ ($\upsilon_{\rm cb}=2\times 10^{-4}\gamma$). To simulate an infinite system, we use periodic boundary conditions by adding repeat images of the system to the boundaries. We truncate to $101$ images along the $\hat{\bf x}$ and $\hat{\bf y}$ direction, which gives an effective lattice size of $406\times 406$, 
and non-zero linewidth for the checkerboard mode.

\begin{figure}
	\hspace*{-0 cm}
	\includegraphics[width=\columnwidth]{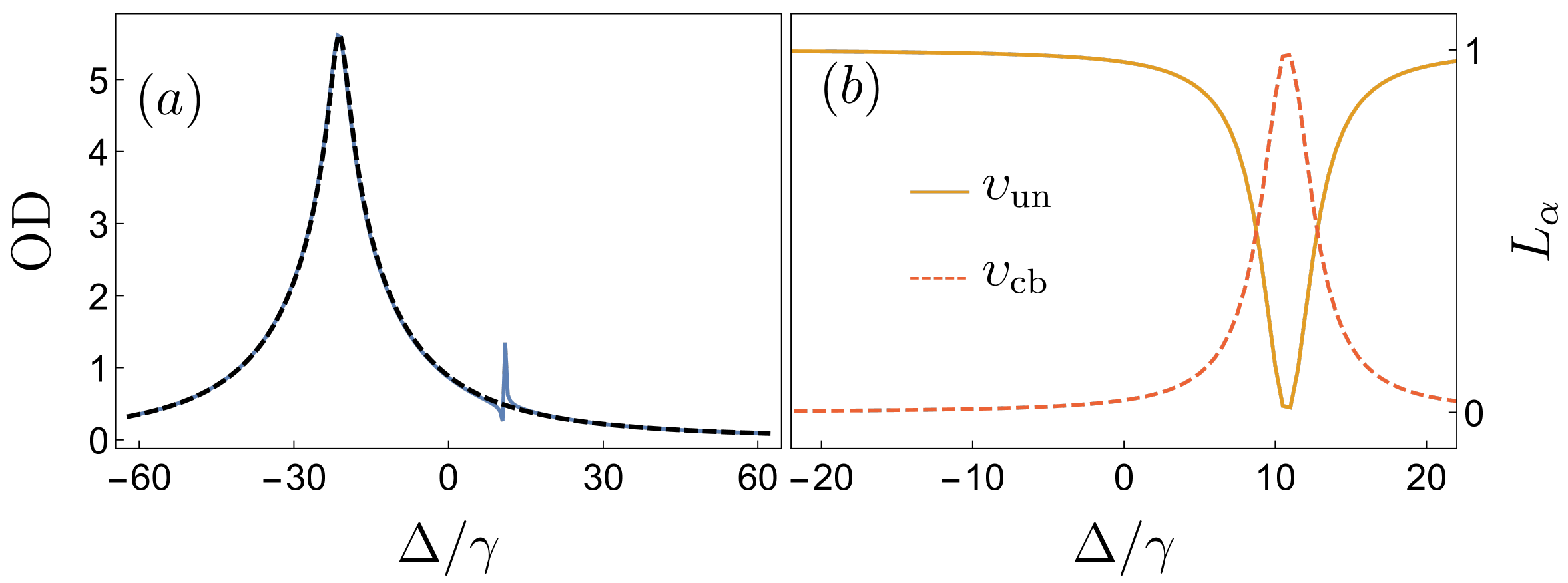}
	\caption{Preparation of the superradiant and subradiant LLI modes by uniform and  checkerboard atomic level shifts. (a) OD for uniform (dashed line) and alternating level shifts of $2\gamma$ (solid line). Alternating level shifts give a Fano resonance between the uniform and checkerboard eigenmodes. (b) The corresponding eigemode populations with alternating level shifts present show the checkerboard subradiant mode strongly populated at the Fano resonance.}
	\vspace{-0cm}
	\label{OpticalResponse}
\end{figure}

The uniform eigenmode $v_{\rm un}$ directly couples to the normally incident light of uniform phase profile, resulting in a broad resonance in the OD, shown in Fig.~\ref{OpticalResponse}(a). We show that we can also excite the checkerboard eigenmode $v_{\rm cb}$ and prepare coherent strongly subradiant excitations that exist outside the light cone. This can be achieved by ac Stark shifts~\cite{gerbier_pra_2006} of lasers (or microwaves) forming a checkerboard pattern of atomic level shifts that results from a standing-wave, $\cos^2[5k(x+y)/\sqrt{2}]$, with the intensity varying along the lattice diagonal $\hat{\bf x}+\hat{\bf y}$ and the intensity maxima separated by $\sqrt{2} a$. 
Alternating blue- and red-detuned atomic transitions for adjacent atoms cause them to oscillate $\pi$ out-of-phase, resulting in the excitation of the checkerboard subradiant eigenmode.
The relative angle between the field generating the ac Stark shift and the lattice can be adjusted to control the periodicity.  
An example of an atomic transition particularly suitable for closely-spaced atoms is $^3P_0\rightarrow ^3D_1$ in $^{88}$Sr~\cite{Olmos13}, which can have
a resonance wavelength of $\lambda\simeq 2.6\mu$m and spacing of 206.4nm, resulting in the effective lattice spacing $a\simeq 0.08\lambda$. 

Figure~\ref{OpticalResponse}(a) shows how these checkerboard-patterned alternating level shifts lead to a coupling to the checkerboard subradiant mode, producing a Fano resonance in the OD at $\Delta=10.8\gamma$, with
the corresponding large population of  the checkerboard eigenmode at this resonance
[Fig.~\ref{OpticalResponse}(b)].

%

%%%%%%%%%%%%%%%%%%%%%%%%%%%%%%%%%%%%%%%%%%%%%%%%%%%%%%%%%%%%%%%
\subsection{Analytic results for optical bistability}

Classifying the steady states of the mean-field solutions of Eqs.~\eqref{SpinEqs} determines the phases that emerge as a function of detuning and incident intensity. 
We calculate the general phase diagram numerically.
However, it is important to understand the collective effects in optical bistability by first deriving solutions in some special cases analytically. 
In order to do so, we consider the uniform case by substituting $\rho_{ge}^{(l)}=\rho_{ge}$, $\rho_{ee}^{(l)}=\rho_{ee}$, and $\Delta_l =\Delta$ into Eqs.~\eqref{SpinEqs}. We then obtain the stationary states 
\begin{subequations}
	\label{GeneralSolns}
	\begin{align}
	\rho_{ge} &={\cal R}_{\rm eff}\frac{-\Delta+\text{i}\gamma}{\Delta^2+\gamma^2+2|{\cal R}_{\rm eff}|^2}, \label{GeneralSolns_1}\\
	\rho_{ee} &=\frac{|{\cal R}_{\rm eff}|^2}{\Delta^2+\gamma^2+2|{\cal R}_{\rm eff}|^2} \label{GeneralSolns_2},
	\end{align}
\end{subequations}
where we have defined 
\begin{equation}\label{E}
{\cal R}_{\rm eff}= {\cal R}+(\tilde\Omega+\text{i}\tilde \gamma)\rho_{ge},
\end{equation}
which, with $\tilde\Omega=\sum_{j\neq l}\Omega_{jl}$ and $\tilde \gamma=\sum_{j\neq l}\gamma_{jl}$,  is the total external electric field (incident plus scattered field from all the other atoms, given in terms of the Rabi frequency) driving an arbitrary atom $l$ in the ensemble.
Eqs.~\eqref{GeneralSolns} are equivalent to the familiar solutions of the independent-atom optical Bloch equations [Eqs.~\eqref{OpticalBlochSolutions} in Appendix C], but with the Rabi frequency ${\cal R}$ replaced by ${\cal R}_{\rm eff}$. As
$\rho_{ge}$ appears on the both sides of Eq.~\eqref{GeneralSolns_1} via ${\cal R}_{\rm eff}$, we generally have multiple solutions. For two different coexisting stable solutions, we have optical bistability.

We can eliminate from Eqs.~\eqref{GeneralSolns_1} and Eq.~\eqref{E} the atomic variables and obtain an equation for the incident light field ${\cal R}={\cal R}({\cal R}_{\rm eff})$. The bistability threshold is then found
when $d |{\cal R}|^2/ d |{\cal R}_{\rm eff}|^2 = 0$. This gives a cubic polynomial for $|{\cal R}_{\rm eff}|^2$ [see Eq.~\eqref{Cubic} in Appendix D] in terms of $\gamma$, $\Delta$, $\tilde\Omega$, and $\tilde\gamma$. Simple analytic expressions for the optical bistability threshold can then
be obtained for $\Delta/\gamma = \tilde\Omega /\tilde{\gamma} $, yielding
$\tilde{\gamma} >8 \gamma$,
and for $\Delta/\gamma = -\tilde{\gamma} /\tilde\Omega $, yielding
$\tilde{\Omega}^2>27 \gamma^2$.
Below these values, there is no bistability for any intensity.
We can also obtain analytic forms for the bistable solutions of the external field ${\cal R}_{\rm eff}$ acting on an atom, for $\tilde{\Omega},~\tilde{\gamma} \gg \Delta^2$, 
\begin{subequations}
	\label{CoopandSingleAtomSoln}
	\begin{align}
	{\cal R}_{\rm eff} &= \frac{\mathcal{R}}{C}\frac{1}{1+\sqrt{1-2 p_{\rm unsat}/|C|^2}}, \label{CoopSoln}\\
	{\cal R}_{\rm eff} &= \frac{\mathcal{R}}{2}\left[1-\frac{2\text{i}\text{Im}[C]}{p_{\rm unsat}}+\sqrt{1-4\frac{\text{Re}[C]}{p_{\rm unsat}}-\left(\frac{2\text{Im}[C]}{p_{\rm unsat}}\right)^2}\right] \label{SingleAtomSoln},
	\end{align}
\end{subequations}
where we have defined the single-atom excited state occupation for unsaturated drive, $p_{\rm unsat} =|\mathcal{R}|^2/( \Delta^2+\gamma^2)$, and the \emph{cooperativity parameter},
\begin{equation}\label{CoopParam}
C  = \frac{1}{2} \frac{\tilde\Omega+\text{i}\tilde{\gamma}}{\Delta+\text{i}\gamma}.
\end{equation}
The two solutions represent very different responses to the incident light. 
The first ``cooperative'' solution, Eq.~\eqref{CoopSoln} (in an analogy with the terminology of optical bistability in cavities~\cite{Bonifacio1978}), exists for $p_{\rm unsat} <  |C|^2/2$ and arises due to the atoms behaving collectively, creating a field that counteracts the incident light and resulting in the atoms absorbing strongly, with enhanced absorption for larger atom density. 
The second ``single-particle'' solution, Eq.~\eqref{SingleAtomSoln}, exists when $p_{\rm unsat} > 2(|C|+\text{Re}[C])$, and arises when the atoms react to the incident light almost independently, with $\mathcal{R} \approx \mathcal{R}_{\rm eff}$ when $|\mathcal{R}| \rightarrow \infty$. The atoms now saturate and absorption is weak, with the medium becoming transparent. 

The simplest system exhibiting collective interactions is that of two atoms ($\tilde\Omega=\Omega_{12}$, $\tilde \gamma=\gamma_{12}$). In this case, we can satisfy $\tilde{\Omega}^2>27 \gamma^2$ for closely spaced atoms for $\Delta/\gamma = -\tilde{\gamma} /\tilde\Omega $. Approximating the resonant dipole-dipole coupling by $\Omega_{12}\sim 1/(ka)^3$, where $a$ denotes the atom separation, results in the bistability threshold of roughly $ka\alt1$, with the precise value depending on the orientation of the dipoles (see Appendix E). 

Analytic expressions can be obtained for atomic chains and arrays  for $\Delta/\gamma =  \tilde\Omega /\tilde{\gamma} $, where the bistability threshold is independent of $\tilde\Omega$.
For an infinite 1D chain, we can sum the series of dissipative dipole-dipole interaction terms over the atoms to obtain the collective resonance linewidth
\begin{equation}
\tilde{\gamma}_{\rm 1D} =\sum_{j\neq l}\gamma_{jl} = \frac{3\gamma\pi}{4 ka}\left[(\hat{\textbf{r}}\cdot \hat{\textbf{e}})^2+1\right]-\gamma,
\end{equation}
where $\hat{\textbf{r}}$ indicates the atomic chain orientation.
The bistability threshold $\tilde{\gamma} >8\gamma$ is met when $k a < \pi/6$ ($ a \alt 0.08\lambda $) or $k a < \pi/12$ ($a \alt 0.04\lambda $) for dipoles parallel and perpendicular to the chain, respectively.
For an infinite 2D array, with a uniform distribution of atomic dipoles in the plane, we obtain  \cite{CAIT} (see also Ref.~\cite{Facchinetti18}) for the collective linewidth
$\tilde{\gamma}_{\rm 2D}/\gamma={3\pi}/{(ka)^2}-1,$
which allows for larger lattice spacings, $k a < (\pi/3)^{1/2}$ ($a \alt 0.16\lambda$), for the bistability threshold than a 1D chain. For dipoles normal to the plane, $\tilde{\gamma}  = -\gamma$, and so $\tilde{\gamma} \not> 8\gamma$ 
and bistability is not possible.

The analogy between the optical bistability in atom arrays and  in cavities~\cite{Gibbs1976,bonifacio1976,Bonifacio1978,Carmichael1977,Agrawal79,Drummond1980,Drummond1981,Savage1988a}
can now be most easily illustrated, and our adapted terminology motivated, at the specific value of $\Delta/\gamma =  \tilde\Omega /\tilde{\gamma} $ for which $C=\tilde{\gamma}/2\gamma$ in Eq.~\eqref{CoopParam} is real.
The expression for the incident light field [Eq.~\eqref{GeneralThreshold2} in Appendix D] then has a similar form as that in cavity systems, with the same 
formulaic dependence on the cooperativity parameter in atom arrays as 
that for optical bistability in cavities~\cite{Bonifacio1978}. In cavity QED,  the cooperativity parameter represents recurrent interactions of an atom with light reflecting between the cavity mirrors. In atom arrays for $\Delta/\gamma =  \tilde\Omega /\tilde{\gamma} $, the bistability condition $C\agt 4$ then translates to the density threshold $k a \sim 1$ -- equivalent to the requirement for the existence of substantial recurrent and correlated light scattering, where the light is scattered more than once by the same atom~\cite{Javanainen2014a,Lee16,Javanainen19}. Moreover, as $\tilde\gamma$ in atom arrays takes the role of the atom-cavity coupling coefficient, the condition $C\agt 1$ then also corresponds to the strong coupling regime of cavity QED.

Numerical solutions of the phase diagram in a planar array agree well with the analytic result of the optical bistability $k a < (\pi/3)^{1/2}$ 
for the uniform phases when driving the superradiant eigenmode, and for $k a\agt 0.44\pi$, only one phase persists and no phase transitions occur. 
When specifically targeting the subradiant eigenmode by using alternating checkerboard-patterned level shifts, optical bistability can only be predicted by numerically solving the equations of motion [specifically Eqs.~\eqref{BP Eqs} in Appendix C], with a much smaller spacing $k a\alt 0.34\pi$ needed for the optical bistability and phase transitions to occur. Bistability also occurs at lower intensities, with bistability in the range 
$0.07 \alt I/I_{\rm sat} \alt190$ when driving the subradiant mode compared to $2 \alt I/I_{\rm sat} \alt 406 $ when driving the superradiant eigenmode. 

\begin{figure}
	\hspace*{0 cm}
	\includegraphics[width=\widthscale\columnwidth,clip,angle=0]{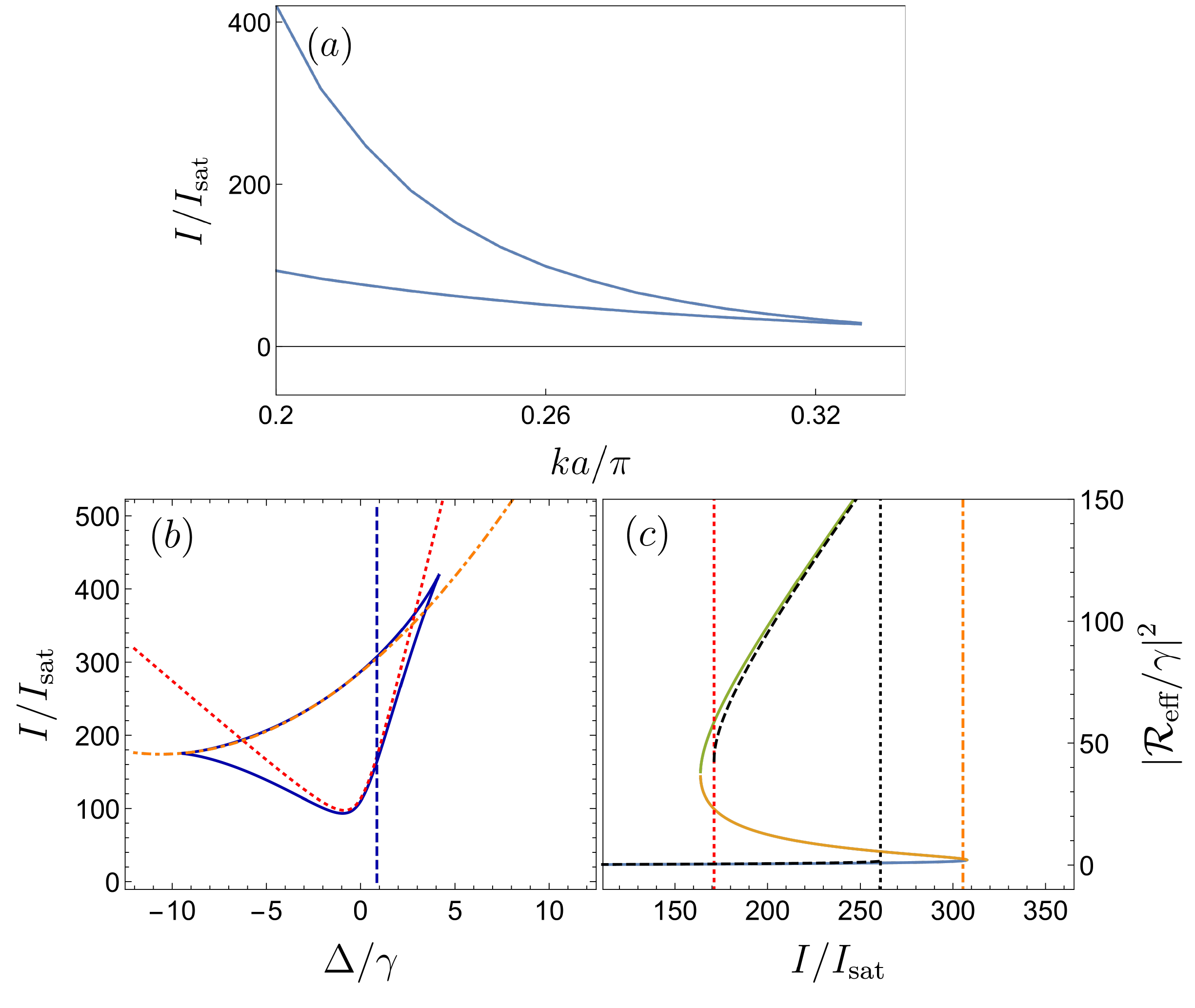}
	\vspace{0 cm}
	\caption{(a) The maximum and minimum intensities of the bistability region for a driven superradiant mode. Both decrease with increasing lattice spacing, with bistability lost for $a \gtrsim 0.16\lambda $ which agrees well with our analytic estimate.
	(b) The region of bistability [from Eq.~\eqref{GeneralThreshold2} in Appendix D] as a function of incident intensity and detuning (solid blue line) for $a=0.1\lambda$; (c) $|{\cal R}_{\rm eff}|^2/\gamma^2$ as a function of incident intensity for $\Delta/\gamma = \tilde \gamma /\tilde\Omega $ [indicated by the blue dashed line in (b)]. In between the lower and upper intensity thresholds, $3$ solutions emerge. The lower (upper) dashed black curve shows the cooperative (single-atom) solution, Eq.~\eqref{CoopSoln} [Eq.~\eqref{SingleAtomSoln}]. In (b,c), the red-dotted and orange-dot-dashed lines show the approximate intensity thresholds [given by Eq.~\eqref{intensitythreshold} in Appendix D], with the analytic estimate for the single-atom solution Eq.~\eqref{SingleAtomSoln}, vanishing at the red-dotted line. In (c), the black-dotted line shows the intensity where the analytic estimate for  the cooperative solution Eq.~\eqref{CoopSoln} vanishes. }
	\label{Bistability}
\end{figure}
%

%
%\begin{figure}
%	\hspace*{0 cm}
%	\includegraphics[width=\widthscale\columnwidth]{images/Figure_1_phasediagramsandmodel.pdf}
%	\caption{A 2D array of atoms illuminated by incident light and resulting stable phases of atoms as a function of laser frequency and incident light intensity. Top: Only the central four atoms (green) are independent, while the remaining array is obtained by periodic boundary conditions. Bottom: (a) Uniform (b) alternating profiles of level shifts.}
%	\vspace{-0cm}
%	\label{qxqyplotBP}
%\end{figure}
%%

%%%%%%%%%%%%%%%%%%%%%%%%%%%%%%%%%%%%%%%%	

%\subsection{Uniform profile of atomic level shifts}
%\subsubsection{Mean-field phase diagram and its optical signatures}

%%%%%%%%%%%%%%%%%%%%%%%%%%%%%%%%%%%%%%%%%%%%%%%%%%%%%%%%%
\subsection{Uniform level shifts: Mean-field phases and optical signatures}
\label{sec:MFobservablesUniform}
So far we have studied the coupling of light in the limit of LLI and the emergence of optical bistability for a uniform atom array.
Next we determine the entire phase diagram of atoms coupled by light-mediated interactions beyond the LLI regime by finding the steady-state solutions of Eqs.~\eqref{SpinEqs} for $k a=0.2\pi$. 
In general, we find the system can exhibit spatially uniform phases, antiferromagnetic (AFM) phases and persistent oscillations (OSC), as well as different optical bistabilities. Transitions between different phases can result in small dips in the OD, and lead to large peaks in the ICR. Phase bistabilities can result in large jumps in the OD and ICR, as well as hysteresis upon varying the laser frequency.

\begin{figure}
	\hspace*{-0 cm}
	\includegraphics[width=\widthscale\columnwidth]{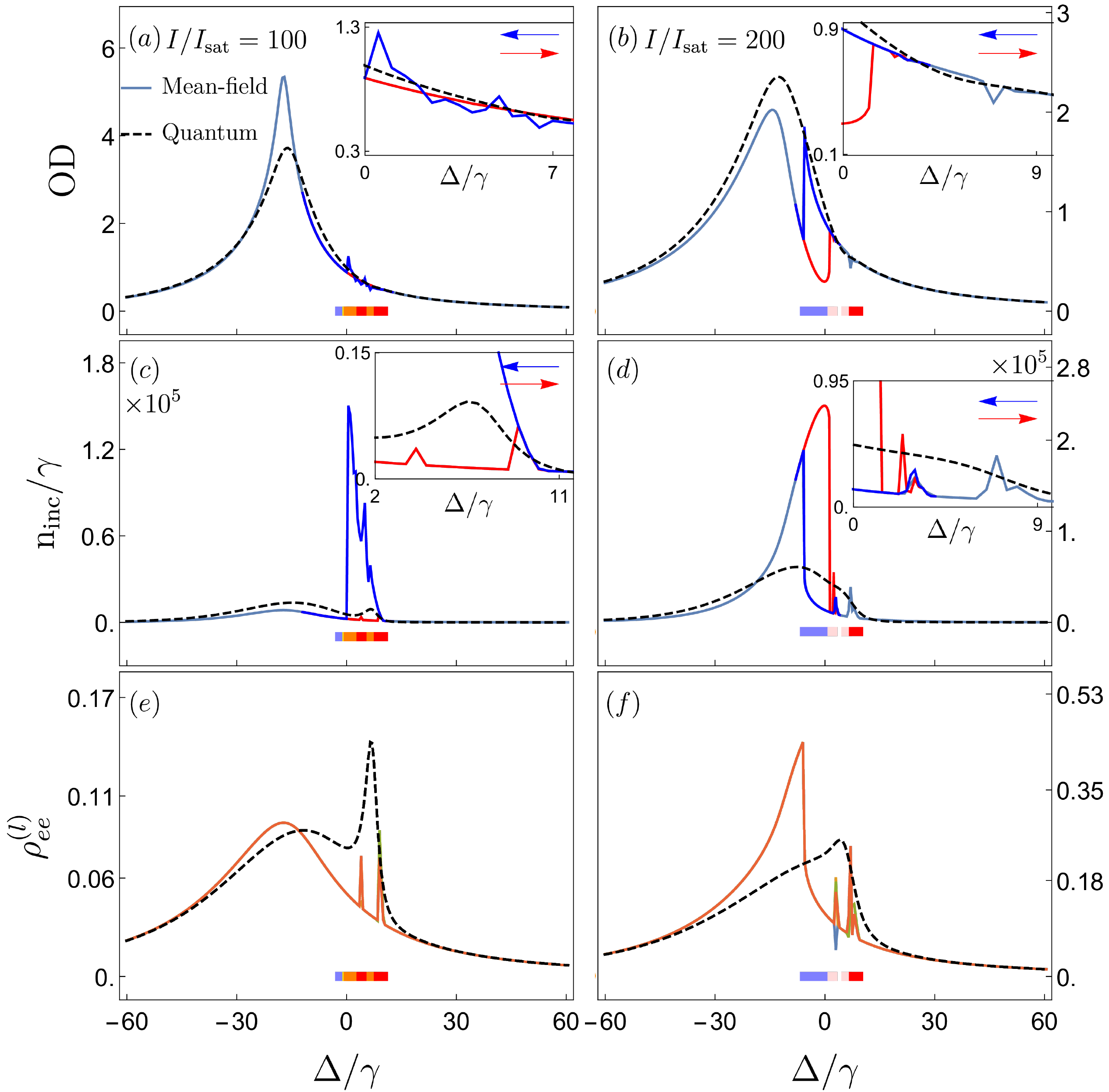}
	\vspace{-0 cm}
	\caption{Signatures of phase transitions in the observables of an array of strongly coupled atoms with uniform level shifts. (a,b) OD, (c,d) ICR and (e,f) number of excitations as a function of detuning for $I/I_{\rm sat}=100$ and $I/I_{\rm sat}=200$. We show hysteresis curves for a negative detuning sweep (red) and positive detuning sweep (blue), with arrows showing the sweep direction. The insets highlight key features on their respective plot. Note we do not show the hysteresis curves for the excitations for clarity. The coloured bars indicate the phases that appear, with the same colouring as used in Fig.~\ref{qxqyplotBP}(a)}
	\label{LowDrivels0.1}
\end{figure}

We first consider the case where the atomic level shifts are all equal; Fig.~\ref{qxqyplotBP}(a). Beyond the obvious phases representing uniform low and high excitation numbers, labelled $\text{U}_1$ and $\text{U}_2$ [the $\textbf{q}=\boldsymbol{0}$ case of Eq.~\eqref{SScoherence} in Appendix C], respectively, we interestingly also find stable phases with spontaneously broken translational symmetries and regions of two coexisting stable phases. The coherent and incoherent optical responses, $\text{OD}$ and ICR, from the mean-field analysis are shown in  Fig.~\ref{LowDrivels0.1}.

While the uniform phases $\text{U}_1$ and $\text{U}_2$ vary smoothly into one another [white regions of Fig.~\ref{qxqyplotBP}(a)], there also exists a $\text{U}_1$/$\text{U}_2$ bistability due to two possible values of $\rho_{ee}$, where the state of the system depends on the initial condition (dark blue region). This bistability region is largely well described by our earlier analytics and contours in Fig.~\ref{Bistability} [derived from Eq.~\eqref{GeneralThreshold2} in Appendix D].
However, differences occur due to one of the uniform phases becoming unstable at positive detunings.
The broad resonances of $\text{U}_{1,2}$ for $I/I_{\rm sat}=100$ in Fig.~\ref{LowDrivels0.1} correspond to the underlying superradiant LLI uniform excitation eigenmode [Eq.~\eqref{uniform} in Appendix B]. For the $\text{U}_1$/$\text{U}_2$ bistability at $I/I_{\rm sat}=200$, we find hysteretic behaviour upon sweeping the resonance from either red- or blue-detuned side. This demonstrates how crossing a region of bistability results in a large jump in both the OD and ICR, with the jump point depending on the initial condition.

Despite the uniformly excited atoms, stable phases with spontaneously broken translational symmetries emerge with the atomic dipoles oscillating $\pi$ out of phase in the neighbouring sites [red regions in Fig.~\ref{qxqyplotBP}(a)].  These AFM phases appear at detunings resonant with the LLI excitation eigenmodes ${\bf u}_{\pm,{\rm cb}}$ [Eqs.~\eqref{striped} and Eq.~\eqref{checkerboard} in Appendix B], and have the same underlying spatial variation, with a striped $\text{AFM}_{\pm}$ phase originating at $\Delta = 4.65\gamma$ and a checkerboard $\text{AFM}_{\text{cb}}$ phase originating at $\Delta = 10.8\gamma$.
The AFM phases materialise as nonlinear interactions between the atoms allow small fluctuations to populate the spatially nonuniform modes in the system, causing phase instabilities.
Two narrow peaks for $I/I_{\rm sat}=100$ in Fig.~\ref{LowDrivels0.1} in the ICR at $\Delta=3.8\gamma$ and $8.7\gamma$ signal spontaneous symmetry breaking and a phase transitioning from $\text{U}_1$ to the $\text{AFM}_{{\pm}}$ and $\text{AFM}_{\text{cb}}$, respectively. No clear signature of this transition can be seen in the OD.
We find that the AFM phase is bistable with the $\text{U}_1$ phase [yellow regions of Fig.~\ref{qxqyplotBP}(a)]. Switching off the incident drive, the uniform phase decays superradiantly, and the AFM phase decays subradiantly. Therefore, $\text{AFM}_{\rm cb}$/$\text{U}_1$ bistability represents an interesting situation where either a superradiant or subradiant phase can be populated depending on the initial condition. 
The hysteresis associated with the bistability could be utilised as a novel method for preparing subradiant excitations, when a steady-state superradiant mode is transformed into a subradiant one by a laser frequency sweep.

The AFM phases also can become unstable (via Hopf bifurcations) resulting in a phase where the atoms continue to oscillate indefinitely, denoted OSC. Such phases appear as additional peaks at $I/I_{\rm sat}=100$ and $I/I_{\rm sat}=200$. Due to oscillations in the OSC phase, the signal is noisy as the stationary state is no longer well defined. There are also regions of OSC and $\text{U}_1$ bistability. For the OSC/$\text{U}_1$ bistability at $I/I_{\rm sat}=100$, clear hysteresis can be seen in the ICR, but hysteresis in the OD is very small. This is due to the alternating out-of-phase dipoles, which, when summed together to calculate the OD, nearly cancel. The ICR always shows clear peaks and hysteresis as it depends on the excitation strength and not the phase. There are small regions (not marked) near this OSC/$\text{U}_1$ bistable region where a new phase emerges which is neither spatially uniform or AFM in nature, where
two dipoles are out of phase to one another, and the two remaining ones in phase with one other.

%

%%%%%%%%%%%%%%%%%%%%%%%%%%%%%%%%%%%%%%%%%%%%%%%%%%%%%%%%%%%%%%
\subsection{Uniform level shifts: Quantum fluctuations}
\label{sec:QObservablesUniform}

In the full quantum theory, there is no bistable behaviour or phase transitions. While generally at high intensities the mean-field and quantum results are in closer agreement~\cite{bettles2019} as the atoms start to scatter more, at intermediate intensities we find in Fig.~\ref{LowDrivels0.1} considerable deviations where the mean-field solutions display bistability. In the full quantum description, due to quantum correlations between different atoms, the ICR no longer represents the excited level population as in the single-atom quantum description that we use to analyse the mean-field dynamics. However, the ICR still shows a resonance near the detunings where mean-field AFM transitions occur.

\begin{figure}
	\hspace*{-0 cm}
	\includegraphics[width=\widthscale\columnwidth]{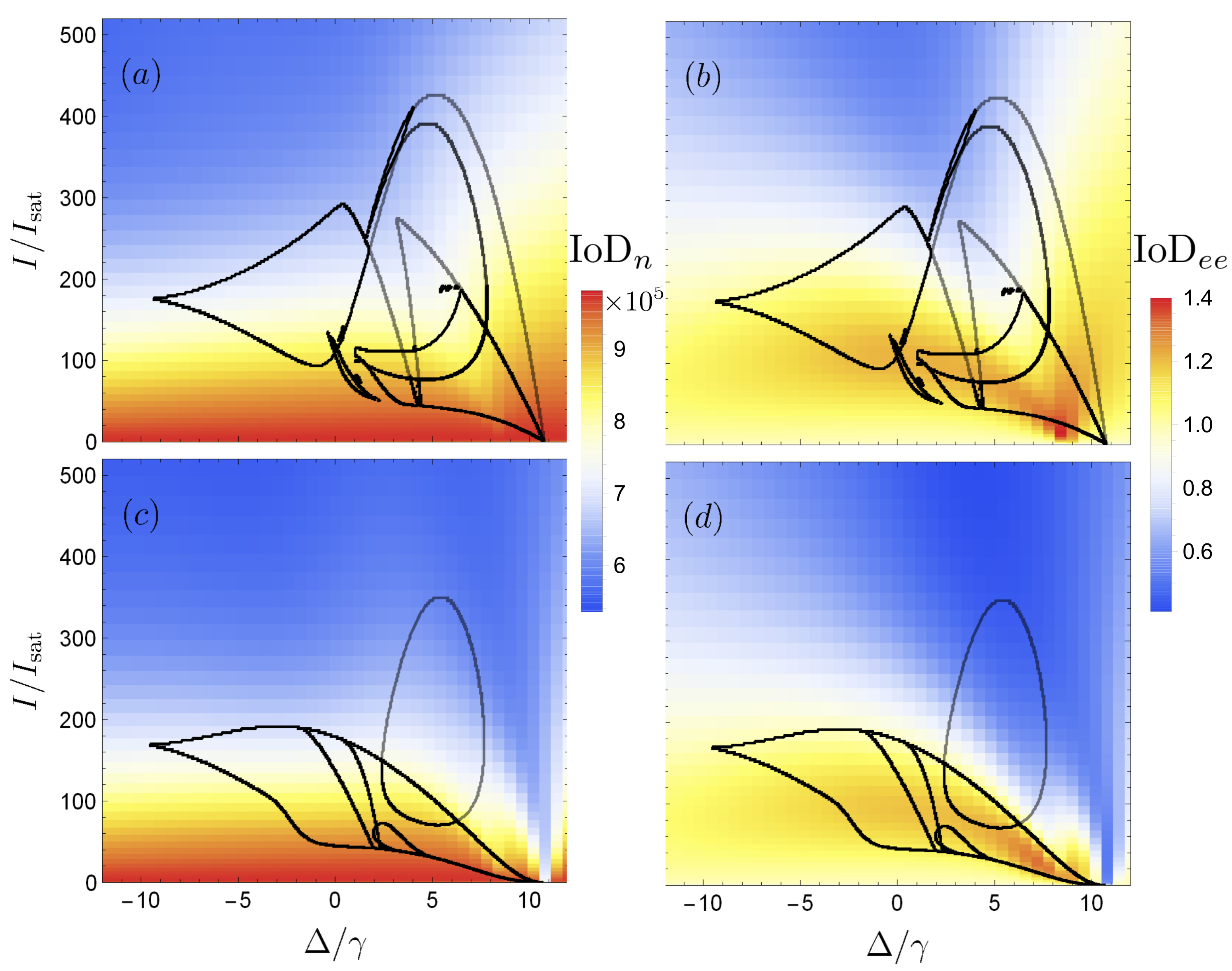}
	\vspace{-0 cm}
	\caption{Quantum theory of incoherently scattered light. Fluctuations of (a,c) scattered photon number (b,d) excited level population with mean-field stability contours (black lines) and mean-field bistability contours (thick black lines) for (a,b) uniform and (c,d) alternating level shift profiles. Fluctuations in excited level population, but not in photon number, are enhanced in regions of mean-field bistability.}
	\label{IoDplot}
\end{figure}

It is generally known from past bistability studies \cite{Carmichael1986a,Olmos2014,Parmee2018,Rempe1991,Drummond1980,Drummond1981} 
that mean-field bistabilities coincide with enhanced quantum fluctuations, which can be understood as tunnelling between the two mean-field solutions. The corresponding quantum distribution is then bimodal. The calculated incoherently scattered photon number fluctuations $\text{IoD}_{n}= (\langle \hat{n}^2\rangle-\langle \hat{n}\rangle^2)/\langle\hat{n}\rangle$, where $\hat{n}$ is the operator form of Eq.~\eqref{TransmissionCoh-2} with all the light collected over a closed surface, however, shows no signatures of enhanced fluctuations [Fig.~\ref{IoDplot}(b)] but we find that the fluctuations of the electronic excitations [Fig.~\ref{IoDplot}(a)],
\begin{equation}\label{IoD}
\begin{split}	
\text{IoD}_{ee}=\frac{\sum_{i,j}\left(\langle \hat{\sigma}_i^{ee}\hat{\sigma}{}_j^{ee}\rangle-\langle \hat{\sigma}{}_i^{ee}\rangle\langle \hat{\sigma}{}_j^{ee}\rangle\right)}{\sum_{i}\langle \hat{\sigma}{}_i^{ee}\rangle},
\end{split}
\end{equation}
are strongly enhanced around the $\text{U}_1/\text{U}_2$ and $\text{AFM}_{\rm cb}/\text{U}_1$ phase bistabilities. This corresponds to large variations in the excitation strength between the different mean-field solutions,
and could be detected by resonantly transferring excited atoms to another level. 

%

%%%%%%%%%%%%%%%%%%%%%%%%%%%%%%%%%%%%%%%%%%%%%%%%%%%%%%%%%%%%%%%%%%%%%%%%
%\subsection{Alternating level shifts}

%\subsubsection{Mean-field phase diagram and its optical signatures}
\subsection{Alternating level shifts: Mean-field phases and optical signatures}
\label{sec:MFobservablesLevelShifts}

\begin{figure}
	\hspace*{0 cm}
	\includegraphics[width=\widthscale\columnwidth]{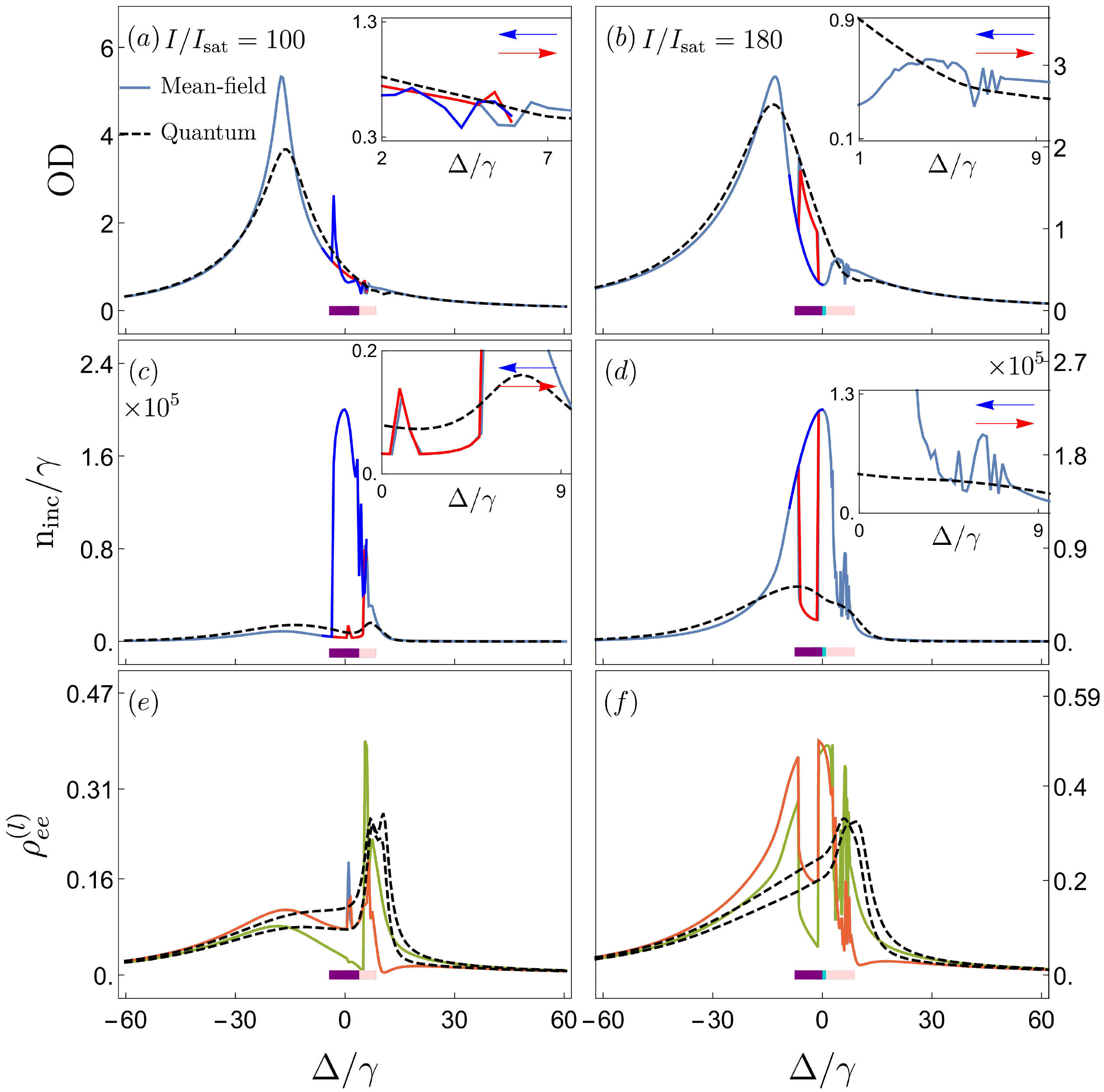}
	\vspace{0 cm}
	\caption{Signatures of phase transitions in the observables of an array of strongly coupled atoms with alternate level shifts. (a,b) OD, (c,d) ICR and (e,f) number of excitations as a function of detuning for $I/I_{\rm sat}=100$ and $I/I_{\rm sat}=180$. We show hysteresis curves for a negative detuning sweep (red) and positive detuning sweep (blue), with arrows showing the sweep direction. The insets highlight key features on their respective plot. Note we do not show the hysteresis curves for the excitations for clarity. The coloured bars indicate the phases that appear, with the same colouring as used in Fig.~\ref{qxqyplotBP}(b)}
	\label{AFMSmallDetuning}
\end{figure}

By engineering a checkerboard pattern of alternating atomic level shifts, detailed earlier in the discussion of the LLI modes,
we are able to drive collective excitations where the atoms oscillate $\pi$ out-of-phase with respect to their nearest-neighbour, and whose LLI limit represents subradiant checkerboard eigenmode [Eq.~\eqref{checkerboard} in Appendix B] existing outside the light cone. We now analyse how this influences the phase diagram beyond the LLI limit.
The alternating level shifts explicitly break the translational symmetry of the lattice. The spatially uniform phases $\text{U}_{1,2}$ of Fig.~\ref{qxqyplotBP}(a) now transform to checkerboard $\text{AFM}_{\rm cb}$ phases in Fig.~\ref{qxqyplotBP}(b), but can still be distinguished as having low and high excitations, labelled $\text{AFM}_{1}$ and $\text{AFM}_{2}$.
AFM phases no longer occur in pairs as the level shifts favour one AFM configuration over the other.
The coherent and incoherent optical responses, $\text{OD}$ and ICR, from the mean-field analysis are shown in  Fig.~\ref{AFMSmallDetuning}. One key difference from the uniform level shifts is that the bistability now occurs at lower intensities,
as discussed earlier when analysing the bistability analytics.
No bistability is found for $I/I_{\rm sat}=200$, so instead we look at $I/I_{\rm sat}=180$. Deviations from the LLI model and the emergence of nonlinear response depend on the linewidth of the corresponding LLI eigenmode, with subradiant modes being more sensitive at lower intensities than superradiant ones \cite{Williamson2020}. Therefore, for non-uniform level shifts, the intensities at which non-trivial phases emerge are lower due to the checkerboard subradiant mode [Eq.~\eqref{checkerboard} in Appendix B] being populated.

There is still a region of OSC phase and also a small region of $\text{AFM}_1$/OSC bistability, indicated by several corresponding peaks in the ICR (and also a dip in the OD for $I/I_{\rm sat}=180$). There are a few cases where the OSC phase becomes unstable and only the $\text{AFM}_1$ phase persists, which are not marked. Subradiant excitations in the limit of LLI lead to narrow Fano resonances when interfering with broader-resonance modes, 
as shown in Fig.~\ref{OpticalResponse}.
Some of these transform to bistable regions, such as $\text{AFM}_1$/$\text{AFM}_2$, which again displays large jumps and hysteresis upon sweeping the detuning. 

Finally, there are regions (green) where phases emerge that are not spatially uniform or AFM in nature
(labelled $\text{BD}_{1,2}$), and are bistable with an AFM phase. For the $\text{BD}_1$ ($\text{BD}_2$) phase, the atoms along $\hat{\bf x}+\hat{\bf y}$ are in-phase (out-of-phase), while the atoms along $\hat{\bf x}-\hat{\bf y}$ are out-of-phase (in-phase). Small regions of the $\text{BD}_2$ phase were found for the uniform shift case. 
Within the $\text{AFM}$/$\text{BD}_2$ bistability region, the $\text{BD}_2$ phase can become unstable and only the AFM phase remains.  
$\text{BD}_{1,2}$ phases occur because of the striped subradiant modes [Eqs.~\eqref{striped} in Appendix B], which have a spatial variation that does not match with the level shift profile, but are populated by nonlinear interactions even though they do not couple to the drive.
There is a small peak in the ICR at $\Delta = 0.9\gamma$ in Fig.~\ref{qxqyplotBP}(b) for the $\text{AFM}_1$ to $\text{BD}_1$ transition, which is similar to the $\text{U}_1$ to $\text{AFM}_{\pm}$ transition peak of Fig.~\ref{qxqyplotBP}(a). 

%

%%%%%%%%%%%%%%%%%%%%%%%%%%%%%%%%%%%%%%%%%%%%%%%%%%%%%%%%%%%%%%%%%%%%%%%%%%%%%%%%
\subsection{Alternating level shifts: Quantum fluctuations}
\label{sec:QobservablesLevelShifts}

The effects of the quantum treatment are similar to those found for uniform level shifts. The quantum system now always exhibits an AFM phase as the alternating level shifts explicitly break the translational symmetry.
The ICR for the quantum model shows peaks around the mean-field phase transitions and we again find enhanced fluctuations in the excitation number around regions of bistability, but not in the photon number; Fig.~\ref{IoDplot}. The enhanced fluctuations appear to agree much better with the mean-field contours, especially around the resonance of the checkerboard mode. Interestingly, there is a large decrease in fluctuations around the resonance of the checkerboard subradiant mode.

%%%%%%%%%%%%%%%%%%%%%%%%%%%%%%%%%%%%%%%%%%%%%%%%%%%%%%%%%%%%%%%%%%%%%%%%%%%	
\section{Discussion}
\label{sec:discussion}

In the limit of LLI, two-level atoms respond to light as linear classical oscillators~\cite{Javanainen1999a}. Although atom-by-atom simulations of such systems, especially in large randomly-distributed ensembles with light-induced positions correlations
between the atoms, can be demanding on numerical resources~\cite{Javanainen2014a}, the number of equations scales linearly with the atom number. Finding full quantum solutions in large systems, however, becomes 
quickly prohibitively challenging as the size of the density matrix in Eq.~\eqref{MasterEq} scales exponentially with the atom number $\sim 2^{2N}$. In this paper, we have approximated the quantum dynamics of a large array
by subjecting it to periodic boundary conditions and varying parameters of only four atoms. This approach, however, provides a useful comparison with the corresponding mean-field dynamics of Eqs.~\eqref{SpinEqs} by
unambiguously identifying quantum effects in the differences between the responses of the two cases.

While identifying light-established quantum correlations is interesting on its own right, this leads to practical implications as the number of equations in the mean-field dynamics scales linearly with the atom number.
Determining the limits of validity of mean-field models can therefore provide a range of useful computational tools for the appropriate regimes. There is also a more philosophical point of view: As experiments with
pristine quantum control of small atomic systems with genuine multimode dynamics are improving, the interface between quantum mechanics and classical physics, and the transition to classical physics due to decoherence
or quantum stochastic nonlinear phenomena, is becoming ever more relevant in many-body systems. When a classical system exhibits the most dramatic consequences of nonlinearity, such as phase transitions or bistability, also the most recognisable differences between
the quantum and classical theories arise. Instability in a classical phase transition represents exponentially growing deviations from the unstable solution to a new stable one, and bistability the simultaneous existence of two stable solutions. Quantum mechanics typically cannot favour either of the corresponding solutions, the dynamics are determined by the initial conditions and the evolution can also emerge as a superposition state, resulting in an enhanced fluctuations of measurement observables, as those identified in our study.

%%%%%%%%%%%%%%%%%%%%%%%%%%%%%%%%%%%%%%%%%%%%%%%%%%%%%%%%%%%%%%%%%%%%%%%
\acknowledgments{J.R.\ and C.D.P.\ acknowledge financial support from the UK EPSRC (Grant Nos.\ EP/P026133/1, EP/M013294/1)}
	
%%%%%%%%%%%%%%%%%%%%%%%%%%%%%%%%%%%%%%%%%%%%%%%%%%%%%%%%%%%%%%%%%%%%%%%
\begin{appendices}
	
\section{Model for light-atom coupling}
We express the electrodynamics in the {\it length} gauge, obtained by the Power-Zienau-Wooley transformation \cite{CohenT}, such that $\boldsymbol{\mathcal{E}}^{\pm}(\textbf{r})=\textbf{D}^{\pm}_{F}(\textbf{r})/\epsilon_0$ correspond to the positive and negative frequency components of the electric displacement in free space. 
The many-body quantum master Eq.~\eqref{MasterEq} is then expressed in the rotating-wave approximation in terms of slowly varying field amplitudes and atomic variables, where $\boldsymbol{\mathcal{E}}^+e^{i\omega t} \rightarrow \boldsymbol{\mathcal{E}}^+$ and $\hat{\sigma}{}^{-}_le^{i\omega t} \rightarrow \hat{\sigma}{}^{-}_l$. 
The dipole-dipole interaction term is given by the real and imaginary part of the dipole radiation kernel,
\begin{equation} 
\frac{1}{\hbar \epsilon_0}\textbf{d}_{eg}\cdot \left[\mathsf{G}(\textbf{r}_j-\textbf{r}_l)\textbf{d}_{ge}\right]=\Omega_{jl}+\text{i}\gamma_{jl},
\label{eq:coupling}
\end{equation}
where $\gamma_{jj}=\gamma$ is the single-atom linewidth.
The dipole radiation kernel acting on a dipole located at the origin yields the familiar dipole radiation expression~\cite{Jackson}
\begin{equation}\label{dipolekernel}
\begin{split}
\mathsf{G}(\mathbf{r})\mathbf{d}&=-\frac{\mathbf{d}\delta(\mathbf{r})}{3}+\frac{k^3}{4\pi}\Bigg\{\left(\hat{\mathbf{r}}\times\mathbf{d}\right)\times\hat{\mathbf{r}}\frac{e^{ikr}}{kr}\\
&-\left[3\hat{\mathbf{r}}\left(\hat{\mathbf{r}}\cdot\mathbf{d}\right)-\mathbf{d}\right]\left[\frac{i}{(kr)^2}-\frac{1}{(kr)^3}\right]e^{ikr}\Bigg\},
\end{split}
\end{equation}
with $r= |\textbf{r}|$, $\hat{\mathbf{r}}=\textbf{r}/r$.

%%%%%%%%%%%%%%%%%%%%%%%%%%%%

For the observables, such as transmitted light intensity or the photon count rate, the electric field product can be expanded in terms of incident and scattered fields to give
\begin{equation}\label{ElectricField}
\begin{split}
&\langle\hat{\textbf{E}}{}^-(\textbf{r})\hat{\textbf{E}}{}^+(\textbf{r}')\rangle=\boldsymbol{\mathcal{E}}{}^-(\textbf{r})\boldsymbol{\mathcal{E}}{}^+(\textbf{r}')+\boldsymbol{\mathcal{E}}{}^-(\textbf{r})\langle\hat{\textbf{E}}{}_{s}^+(\textbf{r}')\rangle+\\
&\langle\hat{\textbf{E}}{}_{s}^-(\textbf{r})\rangle\boldsymbol{\mathcal{E}}{}^+(\textbf{r}')+\langle\hat{\textbf{E}}{}_{s}^-(\textbf{r})\rangle\langle\hat{\textbf{E}}{}_{s}^+(\textbf{r}')\rangle+\langle\delta\hat{\textbf{E}}{}_{s}^-(\textbf{r})\delta\hat{\textbf{E}}{}_{s}^+(\textbf{r}')\rangle,
\end{split}
\end{equation}	
where $\hat{\textbf{E}}{}^-\hat{\textbf{E}}{}^+$ is the dyadic product with elements $E_{\alpha}E_{\beta}^*$, with $\alpha$ and $\beta$ denoting the vector components. The first term in Eq.~\eqref{ElectricField} gives the incident intensity, while the next three terms are the coherent scattered light, which remain even in the absence of quantum fluctuations. The last term, $\langle\delta\hat{\textbf{E}}{}_{s}^-(\textbf{r})\delta\hat{\textbf{E}}{}_{s}^+(\textbf{r}')\rangle =\langle\hat{\textbf{E}}{}_{s}^-(\textbf{r})\hat{\textbf{E}}{}_{s}^+(\textbf{r}')\rangle-\langle\hat{\textbf{E}}{}_{s}^-(\textbf{r})\rangle\langle\hat{\textbf{E}}{}_{s}^+(\textbf{r}')\rangle$, is the incoherent scattering, which is light scattered by disorder and quantum fluctuations. Because we consider atoms at fixed positions, the incoherent scattering is determined purely by quantum correlations. We measure the incoherent scattering from the ICR, obtained by substituting $\langle\delta\hat{\textbf{E}}{}_{s}^-(\textbf{r})\delta\hat{\textbf{E}}{}_{s}^+(\textbf{r}')\rangle$ into the photon count rate expression, Eq.~\eqref{TransmissionCoh-2}, which gives
\begin{equation}\label{IncoherentPhotonCountRateFull}
\begin{split}
n_{\rm inc}=&\frac{2\epsilon_0c}{\hbar \omega_0 }\sum_{l,m}^{N}\left(\langle \hat{\sigma}{}_l^{+}\hat{\sigma}{}_m^{-}\rangle-\langle \hat{\sigma}{}_l^{+}\rangle\langle\hat{\sigma}{}_m^{-}\rangle\right)\cross\\
&\int[\mathsf{G}(\textbf{r}-\textbf{r}_l)\textbf{d}_{ge}]\cdot[\mathsf{G}(\textbf{r}-\textbf{r}_m)\textbf{d}_{ge}]^*dS.
\end{split}
\end{equation} 
The ICR that we will use under the mean-field description reads~\cite{bettles2019} 
\begin{equation}\label{IncoherentPhotonCountRate}
\begin{split}
&n_{\rm inc}=\frac{2\epsilon_0c}{\hbar \omega_0 }\sum_{l}^{N}\left(\langle \hat{\sigma}_l^{ee}\rangle-|\langle \hat{\sigma}_l^{-}\rangle|^2\right)\int|\mathsf{G}(\textbf{r}-\textbf{r}_l)\textbf{d}_{ge} |^2dS.
\end{split}
\end{equation}
As noted in the main text, this expression differs from the usual semiclassical description of the incoherent scattering~\cite{meystre1998} (which would vanish for fixed atomic positions) due to the inclusion of $\langle \hat{\sigma}_l^{ee}\rangle$ terms.
When calculating the photon count-rate, we integrate the field over a solid angle with $\text{NA}=\sin\theta=0.24$,
except when looking at photon fluctuations in Fig.~\ref{IoDplot}, where we integrate over a closed surface.

Coherently transmitted light through a finite array can be approximated at a point $(0,0,\xi)$ from the centre of the array  from~\cite{dalibardexp,Javanainen17,Facchinetti18,Javanainen19}
\begin{equation}\label{EFieldSlab}
\begin{split}
\hat{\textbf{E}}{}^+_j(\textbf{r})=\mathcal{E}_0\hat{\textbf{e}}e^{\text{i}k\xi} + \frac{2\text{i}k}{\mathcal{A}\epsilon_0}\sum_{l}^{}[{\bf d}_{ge}-(\hat{\textbf{z}}\cdot{\bf d}_{ge})\hat{\textbf{z}}]e^{\text{i}k(\xi-z_l)}\hat{\sigma}_j,
\end{split}
\end{equation}	
when $\lambda \lesssim \xi \ll \sqrt{\mathcal{A}} $, where $\mathcal{A}$ is the total area of the array and $l$ is summed over all images of the atom $j$ that are included due to the periodic boundary conditions. The total coherent field is then $\hat{\textbf{E}}{}^+(\textbf{r})=\sum_{j}^{4}\hat{\textbf{E}}{}^+_j(\textbf{r})$. We have found numerically this approximation works well. 
\\

\section{Collective  low light intensity eigenmodes}
In the limit of LLI, the coupled-dipole model or the classical linear oscillator model becomes exact for the two-level atoms~\cite{Javanainen1999a,Lee16}. In the periodic lattice system, the LLI collective excitation eigenmodes are obtained by diagonalizing Eq.~\eqref{eq:coupling}, and are the Bloch waves 
\begin{align}
v^{(+)}_{{\bf q}}({\bf r}_l) &=A_{{\bf q}}\cos(\textbf{q}\cdot\textbf{r}_l), \\ 
v^{(-)}_{{\bf q}}({\bf r}_l) & =A_{{\bf q}}\sin(\textbf{q}\cdot\textbf{r}_l),
\end{align}
where $A_{{\bf q}} = \sqrt{2/N}$ except for $A_{{\bf q} = 0, (\pi/a, \pi/a)} = 1/\sqrt{N}$.
The wavevectors $\textbf{q}$ have components $q_{x/y}=2\pi m_{x/ y}/N_{x/y}a$, where $m_{x/y} = 0, 1,..$ to $N_{x/y}/2$ or $(N_{x/y}-1)/2$ for an even or odd number of sites respectively, and $N_{x/y}$ is the number of sites along the $x/y$ direction. The corresponding eigenvalues, $\delta_{{\bf q}}+ i \upsilon_{{\bf q}}$, represent the collective line shifts (from the resonance of an isolated atom) $\delta_{{\bf q}}$ and linewidths $\upsilon_{{\bf q}}$. 

In our system there are four relevant LLI eigenmodes: the spatially uniform mode,
\begin{equation}\label{uniform}
v_{\rm un}({\bf r}_l)\equiv v^{(+)}_{{\bf q}={\bf 0}} ({\bf r}_l)= \frac{1}{2}, 
\end{equation}
the spatially nonuniform modes with striped phase variation along $\hat{\bf x}\pm \hat{\bf y}$, 
\begin{equation}\label{striped}
v_{\pm}({\bf r}_l)\equiv v^{(-)}_{{\bf q}=(\pi/2a,\pm\pi/2a)} ({\bf r}_l)=\frac{1}{\sqrt{2}}\sin\left[\left(\frac{\pi}{2a},\pm\frac{\pi}{2a}\right)\cdot\textbf{r}_l\right], 
\end{equation}
and a checkerboard phase variation, 
\begin{equation}\label{checkerboard}
v_{\rm cb}({\bf r}_l)\equiv v^{(+)}_{{\bf q}=(\pi/a,\pi/a)} ({\bf r}_l)=\frac{1}{2}\cos\left[\left(\frac{\pi}{a},\frac{\pi}{a}\right)\cdot\textbf{r}_l\right]. 
\end{equation}
To simulate an infinite system, we use periodic boundary conditions by adding repeat images of the system to the boundaries. We truncate to $101$ images along the $\hat{\bf x}$ and $\hat{\bf y}$ direction, which gives an effective lattice size of $406\times 406$. Numerically, the linewidths are given by $\upsilon_{\rm uni}=25\gamma$, $\upsilon_{+}=0.09\gamma$, $\upsilon_{-}=0.08\gamma$, and $\upsilon_{\rm cb}=2\times 10^{-4}\gamma$. Due to our image truncation, the two striped modes $\upsilon_{\pm}$ are not degenerate and all $3$ subradiant modes have a nonzero linewidth.
The population of the eigenmodes  [Fig.~\ref{OpticalResponse}(b)] is calculated using the occupation measure defined by~\cite{Facchinetti16} 
\begin{equation}\label{ModePopulation}
\begin{split}	
L_{\alpha}=\frac{\left|\sum_{l}^{N}v_{\alpha}(\textbf{r}_l)\rho_{ge}^{(l)}\right|^2}{\sum_\beta\left|\sum_{l}^{N}v_{\beta}(\textbf{r}_l)\rho_{ge}^{(l)}\right|^2}, \quad \alpha, \beta = \rm un, \pm, \rm cb.
\end{split}
\end{equation}
	
\section{Bipartite Ansatz}

Most of the steady-state solutions to Eqs.~\eqref{SpinEqs} are obtainable using a bipartite lattice ansatz, where the $4$ independent atoms are divided into $2$ sites,  labelled A and B. For spatially uniform level shifts, two divisions of the lattice can be chosen: sites A and B can lie along the $\hat{\bf x}+\hat{\bf y}$ and $\hat{\bf x}-\hat{\bf y}$ directions, respectively, forming a checkerboard variation, or both A and B sites lie along the $\hat{\bf x}$ ($\hat{\bf y}$) direction, alternating with respect to each other in the $\hat{\bf y}$ ($\hat{\bf x}$) direction, forming a striped variation.   
For the nonuniform checkerboard level shift profile, the A and B sites must lie along the $\hat{\bf x}+\hat{\bf y}$ and $\hat{\bf x}-\hat{\bf y}$ directions, respectively, due to the spatial variation of the level shifts. 

The steady-state solutions to Eqs.~\eqref{SpinEqs} under the bipartite ansatz obey
\begin{subequations}\label{BP Eqs}
	\begin{align}
	\begin{split}
	&\left(\text{i}\Delta_{A/B}-\gamma\right) \rho^{A/B}_{ge}-\text{i}(2\rho^{A/B}_{ee}-1){\cal R}_{A/B}-\text{i}(2\rho^{A/B}_{ee}-1)\cross\\
	&\left[(\Omega_{AA}+\text{i}\gamma_{AA})\rho^{A/B}_{ge}+(\Omega_{AB}+\text{i}\gamma_{AB})\rho^{B/A}_{ge}\right]=0,
	\end{split}\\
	\begin{split}
	&-2\gamma\rho^{A/B}_{ee}+2\text{Im}[{\cal R}_{A/B}^*\rho^{A/B}_{ge}]-2\gamma_{AA}|\rho^{A/B}_{ge}|^2-\\
	&2\text{Im}\left[(\Omega_{AB}-\text{i}\gamma_{AB})\rho^{A/B}_{ge}(\rho^{B/A}_{ge})^*\right]=0,
	\end{split}
	\end{align}
\end{subequations}
where the subscript A/B denotes the lattice site group  and we have made use of the symmetry under $ A \leftrightarrow B$. The stability of the multiple solutions of Eqs.~\eqref{BP Eqs} is studied using linear stability analysis and the results are checked by integrating Eqs.~\eqref{SpinEqs} for $N=4$ atoms.
Additionally, we also simulate the dynamics of all $4$ atoms to find the remaining phases that lie outside the bipartite ansatz. 

For a general form ${\cal R}_l={\cal R} e^{i\textbf{q}\cdot\textbf{r}_l}$, where $\textbf{q}$ is the wavevector of the drive, a solution to Eqs.~\eqref{BP Eqs}, and in general, Eqs.~\eqref{SpinEqs}, is given by $\rho_{ge}^{(l)}=\rho_{ge}e^{i\textbf{q}\cdot\textbf{r}_l}$, with 
\begin{equation}\label{SScoherence}
\begin{split}
\rho_{ge}=\frac{\text{i}{\cal R} (2\rho_{ee}-1)}{\text{i}\left[\Delta-(2\rho_{ee}-1)\tilde\Omega(\textbf{q})\right]-\left[\gamma-(2\rho_{ee}-1)\tilde{\gamma}(\textbf{q})\right]},
\end{split}
\end{equation} 
where 
\begin{equation}\label{FourierTransforms}
\begin{split}
&\tilde\Omega(\textbf{q})=\sum_{j\neq l}\Omega_{jl}e^{\text{i}\textbf{q}\cdot\textbf{r}_{j}},\quad\tilde{\gamma}(\textbf{q})=\sum_{j\neq l}\gamma_{jl}e^{\text{i}\textbf{q}\cdot\textbf{r}_{j}},
\end{split}
\end{equation}
are the Fourier transforms of the real and imaginary parts of the dipole kernel, Eq.~\eqref{dipolekernel},
respectively (excluding the self-interaction $j=l$).
The number of excitations $\rho_{ee}$ obeys the following cubic equation
\begin{equation}\label{SScubic}
\begin{split}
&\big[\tilde{\gamma}(\textbf{q})^2+\tilde\Omega(\textbf{q})^2\big](2\rho_{ee}-1)^3+\\
&\big[\tilde{\gamma}(\textbf{q})^2+\tilde\Omega(\textbf{q})^2-2\Delta \tilde\Omega(\textbf{q})-2\gamma \tilde{\gamma}(\textbf{q})\big](2\rho_{ee}-1)^2+\\
&\big[\Delta^2+\gamma^2+2|{\cal R}|^2-2\Delta \tilde\Omega(\textbf{q}) -2\gamma \tilde{\gamma}(\textbf{q})\big](2\rho_{ee}-1)+\\
&\left(\Delta^2+\gamma^2\right)=0.
\end{split}
\end{equation}
We use the solutions to Eq.~\eqref{SScubic} and Eq.~\eqref{SScoherence} to describe the spatially uniform solutions and their bistability in the phase diagram with uniform level shifts. In the absence of incident light, Eq.~\eqref{SScubic} admits only one real solution of $\rho_{ee}=0$. For large intensities, the interaction terms in Eq.~\eqref{SScubic} become negligible and the coherence and number of excitations become identical to the noninteracting solutions to the optical Bloch equations with familiar power-broadened linewidths
\begin{subequations}
	\label{OpticalBlochSolutions}
	\begin{align}
	\rho_{ge} &={\cal R}\frac{-\Delta+\text{i}\gamma}{\Delta^2+\gamma^2(1+I/I_{\rm sat})},\label{OpticalBlochSolutions_1}\\
	\rho_{ee} &=\frac{I}{2I_{\rm  sat}}\frac{\gamma^2}{\Delta^2+\gamma^2(1+I/I_{\rm sat})}, \label{OpticalBlochSolutions_2}
	\end{align}
\end{subequations}
where we have used $2|{\cal R}|^2/\gamma^2=I/I_{\rm sat}$. 

\section{Bistability threshold}
The regimes of bistability between the spatially uniform phases can be predicted analytically for certain detunings. Without the loss of generality in the derivation, we can set in the following ${\bf q}= {\bf 0}$ for the drive. Substituting $\rho_{ge}^{(l)}=\rho_{ge}$ and $\rho_{ee}^{(l)}=\rho_{ee}$ into Eqs.~\eqref{SpinEqs} gives the steady-state solutions Eqs.~\eqref{GeneralSolns} that are equivalent to the solutions of the optical Bloch equations, Eqs.~\eqref{OpticalBlochSolutions}, but with the Rabi frequency ${\cal R}$ replaced by the total external electric field on an atom (the incident field plus the scattered light from all the other atoms) in the array, ${\cal R}_{\rm eff}$ [Eq.~\eqref{E}].
By substituting 
Eq.~\eqref{GeneralSolns_1} into Eq.~\eqref{E}, we obtain 
\begin{equation}\label{GeneralThreshold2}
\begin{split}
\mathcal{R}&={\cal R}_{\rm eff}+{\cal R}_{\rm eff}\frac{2C(\Delta^2+\gamma^2)}{\Delta^2+\gamma^2+2|\mathcal{R}_{\rm eff}|^2},
\end{split}
\end{equation}
where $C$ is given by Eq.~\eqref{CoopParam}
for  $\tilde\Omega \equiv \tilde\Omega(\textbf{0})$ and $\tilde{\gamma} \equiv \tilde{\gamma}(\textbf{0})$. 
Eq.~\eqref{GeneralThreshold2} is also valid for a general plane-wave Rabi drive ($\textbf{q}\neq\textbf{0}$) if $\textbf{0}\rightarrow\textbf{q}$
and $\rho_{ge}\rightarrow\rho_{ge}e^{\text{i}\textbf{q}\cdot\textbf{r}_j}$ in Eq.~\eqref{E}.
Analytic solutions to Eq.~\eqref{GeneralThreshold2} can be found by either ignoring the single ${\cal R}_{\rm eff}$ term, or the $\Delta^2+\gamma^2$ term in the denominator, which gives the cooperative [Eq.~\eqref{CoopSoln}] and single-atom [Eq.~\eqref{SingleAtomSoln}] solutions in the main text, respectively.
It is worth noting that for real $C$, Eq.~\eqref{GeneralThreshold2} has similar form as equations determining bistability in cavities~\cite{Bonifacio1978}, as discussed in the main text. 

The threshold of bistability is determined by 
taking the modulus-squared of Eq.~\eqref{GeneralThreshold2} (and using $I/I_{\rm sat}=2|{\cal R}/\gamma|^2$) and then
finding the values of $|{\cal R}_{\rm eff}|^2$ that minimise $I/I_{\rm sat}$, resulting in a cubic equation,
\begin{equation}\label{Cubic}
\begin{split}
&4|{\cal R}_{\rm eff}|^2\left(\eta^2+2|{\cal R}_{\rm eff}|^2\right)\left(\eta^2+2|{\cal R}_{\rm eff}|^2+\alpha\right)\\
&+\left(\eta^2-2|{\cal R}_{\rm eff}|^2\right)\left[\beta^2+\left(\eta^2+2|{\cal R}_{\rm eff}|^2+\alpha\right)^2\right]=0,
\end{split}
\end{equation}
where $\eta^2=\gamma^2+\Delta^2$ and $\alpha +\text{i}\beta = 2(\Delta^2+\gamma^2)C$.
Bistability occurs when two positive real solutions to Eq.~\eqref{Cubic} are found.
When $\Delta/\gamma = \tilde\Omega /\tilde{\gamma} $, $\beta=0$, and the bistability threshold is
$\tilde{\gamma} >8 \gamma$,
while for $\Delta/\gamma = -\tilde{\gamma} /\tilde\Omega $, $\alpha=0$, and the threshold is
$\tilde{\Omega}^2>27 \gamma^2$.
Below these values, there is no bistability for any intensity at the respective detuning.

In the limit that $\tilde\Omega ,\tilde{\gamma}   \gg \Delta^2+\gamma^2$, the intensity range for bistability is  approximately given by
\begin{equation}\label{intensitythreshold}
\begin{split}
&\frac{2}{\gamma^2}\left(\alpha \pm \sqrt{\alpha^2+\beta^2}\right) <\frac{I}{I_{\rm sat}}<\frac{\left[\left(2\Delta^2+2\gamma^2+\alpha\right)^2+ \beta^2 \right]}{4\gamma^2(\Delta^2+\gamma^2)},
\end{split}
\end{equation}
where the sign of the square root is chosen such that $I/I_{\rm sat}$ is always positive. For $\Delta/\gamma = \tilde\Omega /\tilde \gamma $, this gives an intensity range of
\begin{equation}\label{intensitythresholdRes1}
\begin{split}
(1+\chi)\frac{2\tilde\gamma }{\gamma}\frac{\tilde G ^2}{\tilde\gamma ^2} <\frac{I}{I_{\rm sat}}< \frac{\tilde G ^2}{\tilde\gamma ^2}\left(1+\chi\frac{\tilde\gamma}{\gamma}+\frac{\tilde\gamma^2}{4\gamma^2}\right),
\end{split}
\end{equation}
where $\tilde G ^2= \tilde\Omega ^2+\tilde\gamma ^2$ and $\chi=1$, while for $\Delta/\gamma = -\tilde{\gamma} /\tilde\Omega $, we have Eq.~\eqref{intensitythresholdRes1} with the following interchange of parameters: $\tilde{\gamma} \leftrightarrow \tilde\Omega $
and $\chi=0$.

\section{Two-atom bistability}

\begin{figure}
	\hspace*{0 cm}
	\includegraphics[width=\widthscale\columnwidth]{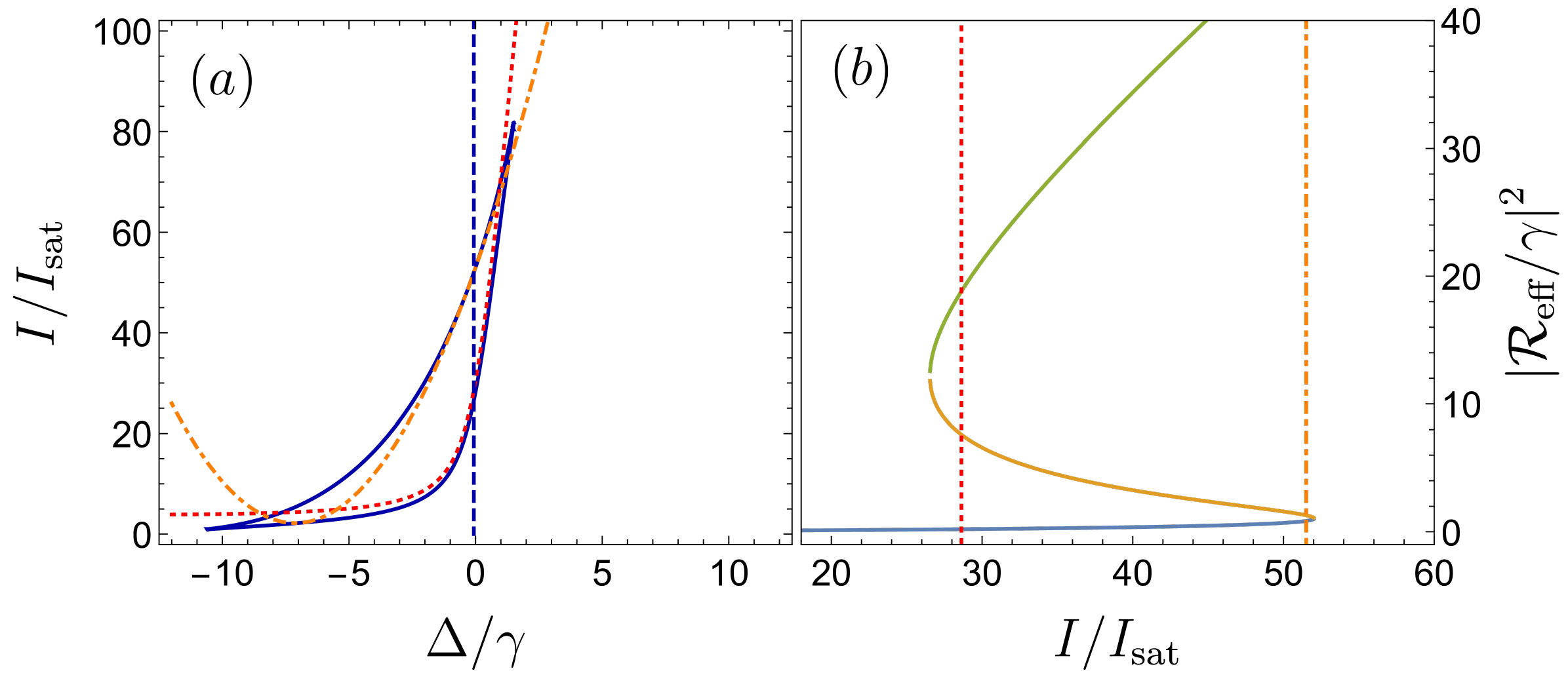}
	\vspace{0 cm}
	\caption{Bistability of two atoms separated by $0.1\lambda$ with the dipoles parallel to the separation axis. (a) The region of bistability as a function of incident intensity and detuning as the solution to Eq.~\eqref{GeneralThreshold2} (solid blue line); (b) $|{\cal R}_{\rm eff}/\gamma|^2$ as a function of incident intensity for $\Delta/\gamma = -\Omega_{12}/\gamma_{12}$ [indicated by the blue dashed line in (a)]. In between the lower and upper intensity thresholds, $3$ solutions emerge. In both plots, the red (dotted) and orange (dot-dashed) lines show the intensity approximation given in Eq.~\eqref{intensitythreshold}.}
	\label{Bistability2Atoms}
\end{figure}

For two atoms, $ \tilde\Omega +\text{i} \tilde{\gamma}  =  \Omega_{12}+\text{i} \gamma_{12}$. Bistability is never possible for $\Delta/\gamma = \Omega_{12}/\gamma_{12}$ as $\gamma_{12}\rightarrow \gamma$ in the limit of close atom spacings, so $\gamma_{12} \ngtr 8\gamma$. For $\Delta/\gamma = -\gamma_{12}/\Omega_{12}$, the threshold is $\Omega_{12}^2 > 27 \gamma^2$, and for close spacing $\Omega_{12}\sim 1/(ka)^3$, such that $ka\alt1$ is generally needed for bistability to emerge. 
Numerically, we find  $a \alt0.15\lambda $ and $a \alt0.10\lambda $ for atoms polarised parallel or perpendicular to the atomic axis, respectively. In Fig.~\ref{Bistability2Atoms}(a), we plot the region of bistability for two atoms showing the intensity limits predicted by Eq.~\eqref{intensitythreshold}. Figure~\ref{Bistability2Atoms}(b) shows the variation of ${\cal R}_{\rm eff}/\gamma$ with $I/I_{\rm sat}$, explicitly showing the expected bistability curve when multiple solutions emerge.

\end{appendices}
	
%\nocite{*}
%\bibliographystyle{apsrev4-1}
%\bibliography{atomlight}

%merlin.mbs apsrev4-1.bst 2010-07-25 4.21a (PWD, AO, DPC) hacked
%Control: key (0)
%Control: author (0) dotless jnrlst
%Control: editor formatted (1) identically to author
%Control: production of article title (0) allowed
%Control: page (1) range
%Control: year (0) verbatim
%Control: production of eprint (0) enabled
%

\end{document}